\documentclass[twocolumn,notitlepage,floats,nofootinbib,amsmath,amssymb,aps,prd]{revtex4-1}

\usepackage{amsmath,amssymb,amsfonts,latexsym,cancel,amsthm}

\usepackage{mathtools}
\usepackage{graphicx}
\usepackage{color}
\usepackage{amsbsy}
\usepackage{hyperref}
\usepackage{tikz}
\usepackage{comment}
\usepackage{seqsplit}

\newcommand{\be}{\begin{equation}}
\newcommand{\beq}{\begin{equation}}
\newcommand{\ee}{\end{equation}}

\def\bea {\begin{eqnarray}}
\def\eea {\end{eqnarray}}

\def\dd{{\rm d}}

\definecolor{newgreen}{rgb}{0.0, 0.75, 0.0}
\definecolor{cadmiumgreen}{rgb}{0.0, 0.42, 0.24}

\begin{document}

\title{Fuzzy-Novae }

\author{Francesco Fazzini} \email{francesco.fazzini@fau.de}

\author{Waleed Sherif} \email{waleed.sherif@fau.de}
\affiliation{Institute for Quantum Gravity, Theoretical Physics III, Department of Physics,  Friedrich-Alexander-Universit\"at Erlangen-N\"urnberg, Staudtstr. 7, 91058 Erlangen, Germany.}

\begin{abstract}
We propose a novel phenomenological model of quantum gravitational collapse inspired by loop quantum gravity that ensures a completely regular spacetime evolution. By incorporating quantum gravitational modifications based on local rather than average energy density, our model resolves both the central singularity and the shell-crossing singularities. Numerical simulations reveal that the interplay between local quantum repulsion and gravitational attraction leads to the formation of a stable, outgoing planckian matter/geometry wave. This mechanism allows for a time-like ejection of the entire stellar mass---a \emph{fuzzy-nova}---which signals the end of macroscopic black holes. By providing a concrete dynamical mechanism for matter to escape a macroscopic trapped region, the model sets a new stage for resolving the information paradox and opens a realistic observational window into quantum gravity.
\end{abstract}

\maketitle

\section{Introduction}

The detection of gravitational waves and the horizon-scale imaging of M87* \cite{EventHorizonTelescope:2019dse} and Sgr A* \cite{EventHorizonTelescope:2022wkp} have firmly established black holes as physical realities. Despite this success, the internal dynamics of these objects remain one of the deepest puzzles in theoretical physics. In general relativity, gravitational collapse inevitably leads to a central singularity \cite{Penrose:1964wq}, signaling the breakdown of the classical theory. While quantum gravity is expected to provide a limiting curvature mechanism to resolve this singularity, the complete spacetime evolution of a collapsing star remains problematic.

A significant, yet often overlooked, challenge in stellar collapse models is the emergence of shell-crossing singularities, where different matter layers intersect \cite{,1985ApJ...290..381H,Szekeres:1995gy}. Although considered weak \cite{Tipler:1977zzb}, these singularities plague most spherically symmetric models, even in presence of radial and tangential pressure \cite{Cafaro:2024lre}, and unlike the central singularity, no standard quantum gravitational mechanism has been proposed to cure them. Furthermore, even models that attempt to regularize the collapse often suffer from mass inflation, Eardley instabilities, or the information paradox \cite{Bobula:2026zlq}.

In this Letter, we propose a phenomenological dust model of inhomogeneous collapse, inspired by Loop Quantum Gravity (LQG) \cite{Rovelli:2004tv,Thiemann:2007pyv}, that ensures a completely regular spacetime evolution. The key novelty lies in the resolution of shell-crossing singularities, which triggers a previously overlooked quantum-gravitational effect: the condensation of the collapsing matter into an outgoing propagating matter/geometry wave, which has to be seen as a collective phenomenon of the underlying dust shells dynamics. As the density approaches the Planck scale, the effective gravitational potential becomes locally repulsive, leading to the formation of a local, dynamical anti-trapped region. The outgoing motion of the matter/geometry wave allows the macroscopic trapped region to disappear within a finite time, without violating causality. The  model challenges the standard Hawking complete evaporation paradigm \cite{Hawking:1975vcx}, potentially resolving its associated information paradox \cite{Page:1993wv}, and leading to a new astrophysical phenomenon termed a \emph{fuzzy-nova}.

Unlike classical white hole geometries, the local dynamical anti-trapped region sustaining the matter-geometry wave emission does not suffer of the Eardley instability \cite{Eardley:1974zz}. 
The model challenges the widespread prejudice that, even within a quantum gravitational framework, matter cannot escape from a macroscopic trapped region via a timelike---and thus physical---path. By providing a concrete dynamical mechanism for this transition, we demonstrate that the quantum-corrected geometry allows for a causal exit of matter, effectively bridging the gap between Planck-scale physics and external observers.

\section{The model}
\label{sec2}

We consider the gravitational collapse of a marginally bound, pressureless dust cloud. In Lemaître-Tolman-Bondi (LTB) coordinates, the metric in the marginally bound case is given by
\begin{equation}
\dd s^2 = -\dd t^2 + [r'(R,t)]^2 \dd R^2 + r(R,t)^2 \dd \Omega^2,
\label{eq:LTB}
\end{equation}
where $r'\equiv \partial_R r$. Classically, the evolution is governed by 
\begin{equation}
\dot{r}^2 = \frac{2Gm(R)}{r}~,
\label{eq:classical_ODE}
\end{equation}
where $\Dot{r}\equiv \partial_t r$, and $m(R)=4 \pi\int_0^R \rho(\Tilde{R},t) r^2 r' \dd \tilde{R} $ is the time-independent Misner-Sharp mass. The dust energy density is recovered via the inverse relation
\begin{equation}
\rho(R,t) = \frac{m'(R)}{4\pi r^2 r'}~.
\label{eq:rho}
\end{equation}
The classical collapse is plagued by two distinct pathologies in the matter region ($m' > 0$): the central singularity as $r\rightarrow0$, and the shell-crossing singularity when $r'\rightarrow 0$. In both cases, the density $\rho$ diverges, marking the breakdown of the Einstein field equations.

To regularize the dynamics, we introduce a phenomenological modification inspired by effective Loop Quantum Cosmology (LQC) \cite{Ashtekar:2006wn}. We modify the classical equation \eqref{eq:classical_ODE} by including a limiting curvature mechanism:
\begin{equation}
{\dot{r}}^2 = \frac{2Gm(R)}{r} \left( 1 - \frac{\rho(R,t)}{\rho_{\text{crit.}}} \right)~,
\label{centrale}
\end{equation}
where $\rho_{\text{crit.}} \approx 0.41 \rho_{\text{Pl.}}$ is the LQC critical density, related to the minimum eigenvalue $\Delta$ of the area operator in LQG by $\rho_{\text{crit.}}=3/(8\pi G\gamma^2\Delta)$, where $\gamma$ is the Barbero-Immirzi parameter. By substituting \eqref{eq:rho} into \eqref{centrale}, the modified evolution is explicitly governed by
\begin{equation}
\dot{r}^2 = \frac{2Gm(R)}{r} \left( 1 - \frac{m'(R)}{4\pi  r^2 r'\rho_{\text{crit.}}} \right)~.
\label{eq:PDE}
\end{equation}
This equation represents a fundamental departure from the classical LTB dynamics. While the classical case \eqref{eq:classical_ODE} is a parametric ODE, the quantum correction couples the evolution of neighbouring shells through $r'$, transforming the system into a non-linear partial differential equation (PDE).

The local quantum correction provides a unified mechanism for singularity resolution.
The quantum correction becomes indeed dominant in two distinct regimes: (i) as $r \rightarrow 0$, where it prevents the central singularity by inducing a bounce in the shell dynamics, analogous to cosmological scenarios, and (ii) as $r'\rightarrow 0$, where generates an effective radial repulsion among nearby shells that prevents shell-crossing. 
The limiting curvature mechanism naturally distinguishes between the central (strong) and shell-crossing (weak) singularities. The effective equation \eqref{eq:PDE} reveals indeed that the bounce velocity is modulated by the specific scaling of the energy density, leading to a weaker quantum response in the shell-crossing regime $(1/r')$ compared to the central bounce $(1/r^2)$. This hierarchical regularization ensures a \emph{soft bounce} where shells are prevented from singular overlapping without undergoing the violent acceleration characteristic of the core bounce, producing, as explicitly shown in sec. \ref{sec4}, a stable outgoing matter/geometry wave.

\section{Numerical Implementation}
We solve Eq.~\eqref{eq:PDE} in a mass-adapted finite-volume formulation. This choice follows the physical structure of pressureless collapse. Each dust layer carries a conserved mass, while its areal radius and physical volume evolve dynamically. The enclosed mass is therefore the natural shell label. In this formulation, mass conservation is exact by construction, and the local density entering the quantum correction is obtained directly from the physical volume occupied by each mass element. This is crucial near the bounce, where the relevant structure is a narrow near-Planckian layer. The limiting-density condition is then resolved locally, without relying on numerical derivatives of a steep mass profile.

The initial data are chosen from smooth, monotonically decreasing density profiles with an asymptotically dilute tail. This avoids an artificial matter-vacuum interface, which would otherwise obstruct the triggering of the local bounce mechanism. At the same time, the tail provides a more realistic representation of the stellar environment than an idealized compact support profile, while remaining gravitationally negligible in the exterior region.

We introduce fixed mass boundaries 
\begin{equation}
    0=m_0<m_1<\cdots <m_N=M~,
\end{equation}
where $M$ is the total gravitational mass. The areal radius of each boundary (or equivalently dust shell) is denoted by $r_i(t)$, and the interval $[m_i, m_{i+1}]$ defines a finite-volume mass cell. For the evolution it is convenient to use
\begin{equation}
    U_i(t):=r_i^3(t)~,
\end{equation}
since the physical volume of a spherical cell is then linear in the evolved variables,
\begin{equation}
    \Delta V_{i+1/2}(t_j)
    =
    \frac{4\pi}{3}\left(U_{i+1}(t_j)-U_i(t_j)\right)~.
\end{equation}
Here $j$ is the discretization label of the time coordinate. The cell-centered density is reconstructed as the finite-volume ratio
\begin{equation}
    \rho_{i+1/2}(t_j)
    =
    \frac{m_{i+1}-m_i}
    {\tfrac{4\pi}{3}\left(U_{i+1}(t_j)-U_i(t_j)\right)}~,
\end{equation}
which is the discrete counterpart of Eq.~\eqref{eq:rho}. In this representation, shell crossing is simply the collapse of a cell volume, $U^j_{i+1}-U^j_i\rightarrow 0$, which dynamically drives the local density toward $\rho_{\mathrm{crit.}}$.

The density $\rho^j_{i}$ entering the velocity of a shell is reconstructed from the adjacent cell densities. In the near-critical regime, we use a conservative reconstruction whereby a boundary is allowed to bounce as soon as one of its neighboring cells reaches the limiting density. This prevents a Planckian pocket from being artificially diluted by averaging. The boundary equations are evolved as
\begin{equation}
    \dot r_i
    =
    \varepsilon_i
    \sqrt{\frac{2Gm_i}{r_i}\chi_i},
    \qquad
    \chi_i
    =
    1-\frac{\rho_i}{\rho_{\mathrm{crit.}}}~,
\end{equation}
or, equivalently,
\begin{equation}
    \dot U_i
    =
    3\varepsilon_i
    \sqrt{2Gm_i\chi_i}\sqrt{U_i}~.
    \label{voleq}
\end{equation}
Here $\varepsilon_i=-1$ labels the collapsing branch and $\varepsilon_i=+1$ the expanding branch. The sign change is implemented locally and shell by shell when the corresponding reconstructed density reaches the limiting value.

The bounce is treated as an event in the evolution. The timestep is chosen adaptively from a CFL-type condition based on the relative motion of neighboring shell boundaries, with fixed lower and upper bounds $\Delta t_{\mathrm{min}}\leq \Delta t\leq \Delta t_{\mathrm{max}}$. The upper bound prevents the evolution from stepping across narrow density peaks, while the lower bound avoids indefinite subdivision at machine precision. The continuum condition $\chi_i=0$ is replaced by the \emph{numerical} condition $\chi_i\leq \chi_b$, with $\chi_b\ll 1$, which triggers the bounce at $\rho_i\geq (1-\chi_b)\rho_{\mathrm{crit.}}$. For the active set $\mathcal{A}$ of boundaries that have not yet bounced, a proposed substep $\delta t\leq \Delta t$ is tested through
\begin{equation}
    \Phi(\delta t)
    =
    \min_{i\in\mathcal{A}}
    \left[
        \min\left(
            \chi_i^{\mathrm{pred}}(\delta t),
            \chi_i^{\mathrm{end}}(\delta t)
        \right)
        -
        \chi_b
    \right],
\end{equation}
where $t_b = t_n + \inf\left\{ \delta t\in[0,\Delta t] \,\middle|\, \Phi(\delta t)\leq 0 \right\}$. If $\Phi(\Delta t)>0$, the step is accepted. If $\Phi(\Delta t)\leq0$, the bracketed event time $t_b$ is localized by bisection up to prescribed absolute and relative tolerances. Events occurring within the same localization window are coalesced and treated as a single Planckian layer. In the simulations reported here, the reconstructed density at a bouncing shell satisfies \emph{at least} $\frac{\rho_i}{\rho_{\mathrm{crit.}}}\geq 0.9999$, equivalently the numerical bounce occurs within \emph{at most} one part in $10^4$ of the continuum limiting density.

At the localized event, no shell position, mass, or density is reset. The configuration is continuous. The only operation is the local change of branch, $\varepsilon_i:-1\mapsto +1$, for every active boundary satisfying the bounce condition. Each boundary is then latched to prevent repeated sign flips caused by roundoff-level oscillations around the limiting-density surface. Since $\dot r_i\propto \sqrt{\chi_i}$, the velocity vanishes at the bounce surface, and the branch change continues the same trajectory through its turning point. The outgoing Planckian layer is therefore not imposed externally. It is generated by the local condition $\rho=\rho_{\mathrm{crit.}}$ and by the subsequent interaction between bounced and still-collapsing shells.

Throughout the evolution we monitor mass conservation, shell ordering, the minimum cell volume, the maximum value of $\rho/\rho_{\mathrm{crit.}}$, and the number and position of bounced shells. The trapped and anti-trapped regions are reconstructed from the same evolved geometry. Although the evolution is performed in LTB coordinates, the figures are displayed in $(r,t)$, or Painlevé-Gullstrand, coordinates by post-processing the evolved shell positions, cell-centered densities, and associated geometrical quantities.

\section{Causal death of a macroscopic black hole}
\label{sec4}
\begin{figure}
    \centering
    \includegraphics[width=1.0\linewidth]{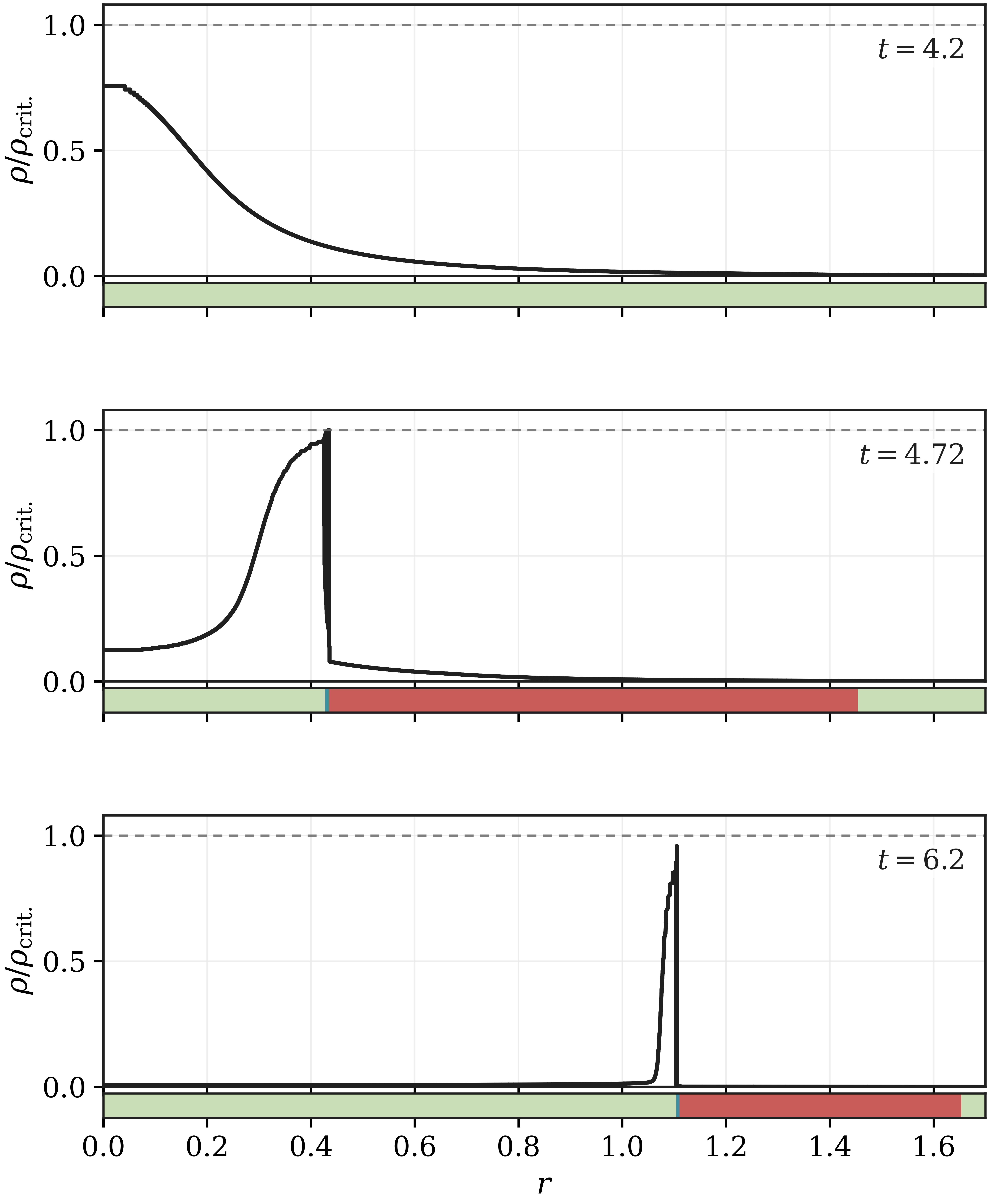}
      \caption[]{\footnotesize Snapshots of the shell dynamics across the bounce for a simulation with a mass $M = 0.85$, $\Delta = 1.0$ and $\gamma=0.2735$. \textbf{Top}: Pre-bounce phase, with the density $\rho$ approaching the critical threshold $\rho_{\text{crit}}$. \textbf{Middle and bottom}: Formation of the matter-geometry wave shortly after the bounce; the wave propagates outward due to local quantum gravitational repulsion. The horizontal bar classifies the spacetime regions: non-trapped (green), trapped (red), and anti-trapped (blue), highlighting the causal exit of matter from the formerly trapped region.}    
      \label{figure}
\end{figure}
Numerical simulations of Eq. \eqref{eq:PDE} reveal a novel dynamical regime.
As the inner core reaches the Planckian density $\rho=\rho_{\text{crit.}}$, quantum gravitational repulsion triggers a bounce. In contrast to standard effective models where shells evolve independently from each other (see e.g. \cite{Husain:2022gwp,Giesel:2023hys,Bojowald:2024ium,Alonso-Bardaji:2023qgu,Fazzini:2026uqp}), the coupling through $r'$ forces the outgoing bounced core layers to interact with the infalling stellar tail. When the density between these disparate layers reaches the Planck scale, the quantum correction triggers a local transition of the geometry from a trapped to an anti-trapped state, passing through a non-trapped state, which in turn allows the matter to move outward.

To characterize this geometry analytically, we examine the null expansions. For the metric \eqref{eq:LTB}, with EoM given by \eqref{centrale} we find
\begin{equation}
\theta_{\pm} = \frac{2}{r} \left[ \pm 1 + \dot{r} \right] = \frac{2}{r} \left[ \pm 1 + \varepsilon \sqrt{\frac{2Gm(R)}{r} \left( 1 - \frac{\rho}{\rho_{\text{crit.}}} \right)} \right]~,
\label{eq:expansions}
\end{equation}
\begin{figure}
    \centering
    \includegraphics[width=1.0\linewidth]{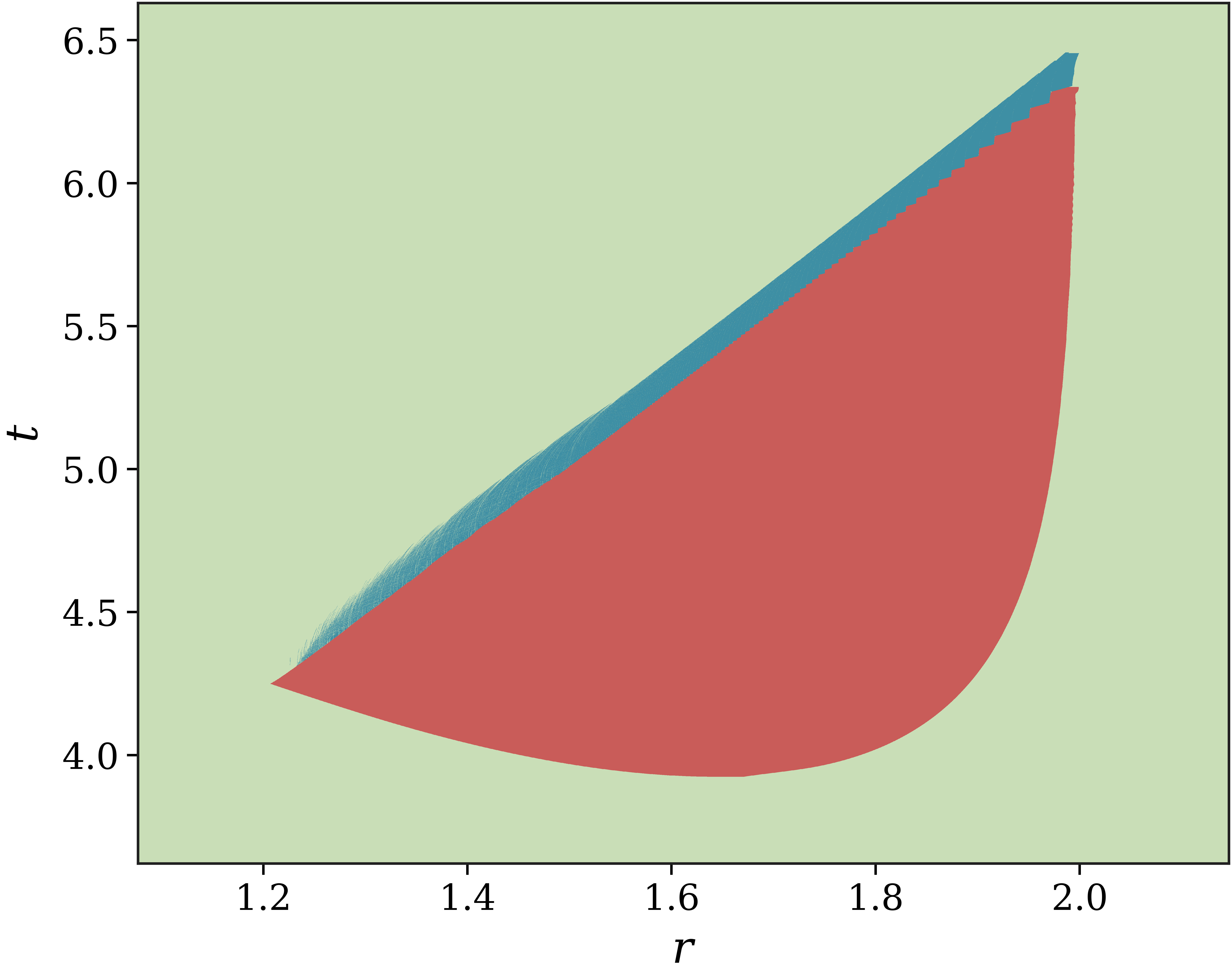}
      \caption[]{\footnotesize Spacetime diagram illustrating the complete gravitational collapse and subsequent rebound for a simulation with a mass $M = 1.0$, $\Delta = 1.0$ and $\gamma=1.0$. A dynamical trapped region (red), bounded by apparent horizons, forms and persists until the conclusion of the process. The matter-geometry wave (traceable along the sharp left boundary of the red region) causal egresses from the trapped interior. The plot highlights the crucial role of the local geometry at the Planckian frontier, where an anti-trapped (blue) region forms at the boundary, effectively tunnelling matter outward in a causal way until the trapped region entirely disappears.}    
    \label{figure1}
\end{figure}
where $\varepsilon=-1$ during the collapse and $\varepsilon=1$ after the local bounce. In the classical regime ($\rho \ll \rho_{\text{crit.}}$), $\theta_\pm <0$ inside the Schwarzschild radius, defining a trapped region. However, at the Planckian threshold, the term in the square root vanishes, leading to a momentarily non-trapped state ($\theta_+ >0$, $\theta_- <0$). Crucially, in the local post-bounce phase, a region emerges where both $\theta_+$ and $\theta_-$ are positive inside the horizon, signifying a local dynamical anti-trapped geometry (see Fig.~\ref{figure1}).

This dynamical local non-trapped region, followed by a dynamical local anti-trapped region---that we name \emph{white wave}---, propagates outward (see Fig.~\ref{figure1}), forming a matter/geometry wave, and its trajectory consists in the space-time points where the dust shells of the stellar tail bounce. The local anti-trapped region exists only as long as the matter density remains near-critical in the local post-bounce phase. Once the matter/geometry wave reaches the horizon and the trapped region disappears, the anti-trapped condition also terminates as $\theta_-$ can only attain negative values for $r>2Gm(R)$.

We have investigated the evolution across a variety of specific initial density profiles, finding that the qualitative dynamics remains remarkably universal. This consistency indicates that the formation of the matter/geometry wave is a generic features of the model, rather than artifacts of a particular initial configuration.

A fundamental requirement for any viable classical or quantum gravitational model of stellar collapse is the preservation of subluminality. We verify this by analysing the light-cone structure. From the null condition $\dd s^2=0$, the radial velocity of outgoing photons on the expanding matter/geometry wave is:
\begin{equation}
\Dot{r}_{\text{ph.}} = \dot{r}_{\text{dust}} + 1~,
\label{eq:photon_vel}
\end{equation}
where $\Dot{r}_{\text{dust}}=\sqrt{\frac{2Gm(R)}{r} \left( 1 - \frac{\rho}{\rho_\text{crit.}} \right)}$ is the velocity of the dust shell located in the areal radius $r$ in the local post-bounce phase. In such a phase the photon radial velocity is positive and strictly greater than the matter velocity. Thus, the matter/geometry wave can be interpreted as a collective wave generated by dust shells each moving in a time-like fashion. It is crucial to note that the wave propagation does not coincide with the underlying dust shell dynamics, and its velocity can be superluminal without violating causality. Such velocity has to be interpreted as a local phase velocity of the stellar fluid, since as we mentioned previously the matter/geometry wave trajectory is given by the spacetime bouncing points of subsequent dust shells. Oppositely, the shells velocities determine the local group velocity of the stellar fluid, which by causality is constrained to be subluminal. To give a clear comparison, in the homogeneous spherically symmetric case, this phase velocity is infinite (since the shells bounce all at the same comoving time).

The quantum gravitational repulsion effectively flips the local light-cone orientation, as shown in Eq. \eqref{eq:photon_vel}. Inside the classical trapped region, all light cones point toward the center. The local bounce generating the white wave locally rotates these cones, allowing an outgoing flux of matter and information to exit the black hole without requiring superluminal propagation or metric discontinuities. This mechanism ensures that the information contained within the stellar layers becomes accessible to an observer lying outside the horizon---as the matter/geometry wave traverses the horizon and the trapped region disappears---, potentially leading to a singularity-free resolution to the information paradox.

Notably, the vacuum sector $m'=0$ of the model remains formally classical. This means that the local bounce mechanism cannot be activated in vacuum, specifically for initial density profiles of compact support. However, in any realistic astrophysical scenario, a perfect vacuum is never realized; the non-vanishing energy density of quantum fields or the presence of even an infinitesimal classical matter tail in the density profile is sufficient to trigger the non-perturbative quantum regime during the collapse, leading to the matter-geometry wave emission. Our model thus suggests that the limiting curvature mechanism is robustly activated by any physical source. 

The transition of the matter-geometry wave across the black hole horizon marks the final dissolution of the black hole. To an external observer, this event appears as the emission of a Planckian matter-geometry wave---which we term fuzzy-nova due to its quantum gravitational nature---followed by a nearly homogeneous, sub-Planckian matter core. Crucially, during the high-density phase, the medium is opaque to electromagnetic radiation, such as during the Planck era in early-universe cosmology. Consequently, the initial stage of the explosion remains dark to an exterior observer, who perceives a transient enlargement of the horizon. As the energy density relaxes to sub-Planckian values, the mean free path of photons increases, rendering the burst visible.

\section{Avoidance of paradoxes and instabilities}

A key feature of this framework is its immunity to the instabilities that plague most regular black hole/stellar collapse models. As established in Sec.~\ref{sec2}, the vacuum sector of the theory is identically Schwarzschild. This precludes the formation of an inner (Cauchy) horizon in the vacuum region, thereby bypassing the mass-inflation instability characteristic of Reissner-Nordstr\"om-like geometries and also effective models inspired by loop quantum gravity \cite{Cao:2023aco}. Furthermore, the Eardley instability \cite{Eardley:1974zz} is avoided because our model does not rely on a global white-hole horizon. Instead, the white wave is a localized, matter-supported dynamical region which is stable against ingoing dust perturbations, as numerically checked.

The absence of physical singularities in our model prevents the destruction of information carried by the collapsing matter. Consequently, the expulsion of the total stellar mass renders this information accessible to external observers. However, to fully address the information paradox \cite{Page:1993wv}, a more stringent condition must be satisfied: the black hole lifetime $T_{\text{BH}}$ must be significantly shorter than the Page time, $T_{\text{Page}} \sim M^3 / m_{\text{Pl}}^2$, the timescale at which half of the Hawking evaporation would have occurred. At the present stage, our numerical analysis does not provide a definitive scaling for $T_{\text{BH}}(M)$. The accumulation of numerical errors, which grows exponentially with the stellar mass $M$, makes the reconstruction of the full $T_{\text{BH}}(M)$ curve---and the subsequent verification of its relation to the Page time---unattainable with the current status of the numerical implementation. Future work will be dedicated specifically to optimizing numerical precision and exploring the large-$M$ regime to confirm whether the fuzzy-nova represents a universal and dynamical resolution to the information paradox.

\section{Comparison with previous models}

The resolution of the central singularity is a shared feature of several effective models inspired by LQG. A prominent example is the model based on gravitational holonomies involving the angular components of the extrinsic curvature in the $\bar{\mu}$-scheme \cite{Husain:2022gwp}. In LTB coordinates, the effective equation for this model reads \cite{Giesel:2023hys}
\begin{equation}
    {\dot{r}}^2 = \frac{2Gm(R)}{r} \left( 1 - \frac{2Gm(R) \gamma^2 \Delta}{r^3} \right),
    \label{eq:old_effective}
\end{equation}
which, by introducing the average energy density $\bar{\rho} = 3m(R)/(4\pi r^3)$, can be recast as:
\begin{equation}
    {\dot{r}}^2 = \frac{2Gm(R)}{r} \left( 1 - \frac{\bar{\rho}}{\rho_{\text{crit.}}} \right).
    \label{eq:old_rho_avg}
\end{equation}
While Eq.~\eqref{eq:old_rho_avg} reduces to standard effective LQC in the homogeneous limit (where $\rho(R,t) = \bar{\rho}(R,t)$), its inhomogeneous evolution is fundamentally limited. Numerical \cite{Husain:2022gwp,Cipriani:2024nhx} and analytical studies \cite{Fazzini:2023ova} have shown that solutions to Eq.~\eqref{eq:old_rho_avg} generically develop shell-crossing singularities within a Planckian time after the bounce. 

The pathology arises because the quantum correction in Eq.~\eqref{eq:old_rho_avg} is triggered by the \emph{average} density $\bar{\rho}$ rather than the local density $\rho$. Consequently, the local energy density can diverge at shell-crossings provided the average density remains bounded. Formally, Eq.~\eqref{eq:old_rho_avg} remains a parametric ODE where each $R$-shell evolves independently, oblivious to the dynamics of neighboring shells. This independence is precisely what allows shells to intersect, rendering the LTB dynamics untrustable beyond the first intersection point and requiring physically consistent spacetime extensions \cite{Bobula:2026zlq}. 

Other models, such as the one inspired by Thiemann cosmology \cite{Assanioussi:2019iye} or by purely phenomenological regular metrics like Hayward and Bardeen geometries \cite{PhysRevLett.96.031103,1968qtr..conf...87B}, similarly fail to provide a limiting curvature mechanism for shell-crossing singularities \cite{Giesel:2026pjj,Fazzini:2026uqp}. Our model departs from the aforementioned by imposing that quantum effects must be governed by a \emph{local} limiting curvature mechanism. As demonstrated in this work, replacing the average density with the local density $\rho(R,t)$ transforms the dynamics into a non-linear PDE. This coupling introduces an effective radial repulsion that regularizes the spacetime evolution at both the center and the matter interfaces.

From a conceptual perspective, our model stands in stark contrast to the standard \textit{black-to-white hole} transition scenario and its associated phenomenology \cite{Rovelli:2024sjl}. The latter typically assumes that quantum gravitational effects remain confined to the deep interior until the black hole reaches a Planckian size, at which point a quantum vacuum transition occurs. We trace this limitation to the idealized assumptions usually adopted in literature, such as the exact Oppenheimer-Snyder collapse, and the systematic non-inclusion of local, dynamical quantum gravitational effects \cite{Han:2023wxg}. Once these ansätze are relaxed, the local bounce mechanism allows for a causal exit of matter from a \emph{macroscopic} black hole. This implies that the resolution of the singularity and the subsequent information release are not delayed until the end of the evaporation process. The stellar content is ejected thanks to the outgoing propagation of a dynamical, matter-geometry wave that progressively reduces the trapped region. Such a wave is a direct consequence of having assumed a local Planck-density/curvature threshold, which in turn is expected by any realistic quantum gravitational model.

\section{On the Validity of the Effective description}
Effective equations in LQG are expected to hold when the physical volume $V \gg \ell_{\text{Pl}}^3$, ensuring that quantum fluctuations are suppressed by a large number of underlying degrees of freedom \cite{Rovelli:2013zaa}. During the pre-bounce phase, the core reaches a minimum radius $r_{\text{min.}} \sim (2G M \gamma^2 \Delta)^{1/3}$ at $\rho_{\text{crit}}$, where $M$ is the total stellar gravitational mass; for $M \gg m_{\text{Pl}} \sim 10^{-8} kg$, the core volume remains many orders of magnitude above the Planck scale, justifying the LQC-like effective treatment. 

In the subsequent matter-geometry wave phase, the wave, located at areal radius $r(t)$, maintains a macroscopically large physical volume $V(t) \approx 4\pi r(t)^2 \delta r \gg \ell_{\text{Pl}}^3$ throughout the ejection process, even for a Planckian radial thickness $\delta r \sim \mathcal{O}(\ell_{\text{Pl}})$. However, while the transverse angular directions are well-sampled by spin networks, the matter-geometry wave thickness approaches the limit where radial fluctuations may become quantitatively significant. Nonetheless, the absence of shocks and singularities suggests that the effective description correctly captures the leading-order regularization mechanism. Further investigations within the canonical framework, group field theory (GFT) or spin foams will be required to resolve the fine-grained fuzziness and potential dispersive effects of the outgoing wave beyond the continuous PDE approximation.

It is worth mentioning that is possible that the inclusion of internal pressure could sufficiently spread the wave, providing the matter-geometry pulse with a macroscopic radial width. Such a mechanism would allow the dynamics to remain within the regime of validity of the effective description, potentially avoiding the need to invoke a full quantum gravitational treatment for the propagation phase.
\section{Conclusions}
In this Letter, we have presented a novel model for quantum gravitational stellar collapse that ensures spacetime regularity throughout the entire dynamical evolution. By implementing a limiting curvature mechanism based on local energy density, we have successfully resolved both the central and shell-crossing singularities, the latter being a persistent pathology in previous effective models. 

Our results demonstrate that the resolution of shell-crossing singularities is not merely a technical refinement, but the catalyst for a new physical phenomenon: the emergence of an outgoing matter-geometry wave. The quantum gravitational correction needed to resolve shell-crossing singularities introduces a new type of local quantum gravitational bounce that allows the matter to move outward in a time-like fashion, sustained by a local dynamical anti-trapped region. This mechanism allows for the complete timelike ejection of the collapsing matter, providing a potential solution to the black hole information paradox without violating causality or triggering known instabilities such as mass inflation or the Eardley instability.

Future work will focus on relaxing the dust approximation by including physical pressure and exploring the multi-messenger signatures expected from the matter-geometry wave ejection. 
Such signals, emitted when the density of the matter-geometry wave becomes sub-Planckian---i.e after the black hole death---may soon be within the reach of next-generation astrophysical observatories, or might even be hidden within existing datasets of unidentified transient phenomena. 

From the theoretical side, future work will aim to derive the effective equations of motion presented here from a modified canonical theory and within the GFT framework. Furthermore, the local bounce dynamics that leads to the propagation of the matter-geometry wave could be rigorously investigated through spin foam computations, providing a covariant path-integral validation of the effective model. This theoretical bridge would not only clarify the microscopic origin of the singularity resolution but also establish the observed dynamics of black holes as a direct, macroscopic probe of the fundamental quantum nature of spacetime.

\acknowledgments
F.F. wishes to thank his wife, Giulia, for her love and continuous support, and for having given birth to the brightest nova of his universe, Chiara. F.F. wants also to thank Carlo Rovelli, Edward Wilson-Ewing and Lorenzo Cipriani for useful discussions. The authors gratefully acknowledge the scientific support and HPC resources provided by the Erlangen National High Performance Computing Center (NHR@FAU) of the Friedrich-Alexander-Universität Erlangen-Nürnberg (FAU). The hardware is funded by the German Research Foundation (DFG).

\bibliographystyle{bib-style}
\bibliography{BibliographyList}

\end{document}